\documentclass[preprint]{sigplanconf}

\usepackage{path}
\usepackage{url}
\usepackage{amsmath,amssymb} 
\usepackage{fancyvrb}

\newcommand{\useverb}[1]{\begin{small}\BUseVerbatim{#1}\end{small}}
\newcommand{\codefig}[2]{\begin{figure}[h]\useverb{#1}\caption{#2}\label{#1}\end{figure}}
\usepackage{graphics,epsfig}
\usepackage{stmaryrd}

\newcounter{per}
\newcommand{\LamPi}{$\lambda_\Pi$}

\begin{document}

\conferenceinfo{POPL 2016}{Jan, 2016}
\copyrightyear{2016}

\title{ Towards a Direct, By-Need Evaluator for Dependently Typed Languages}

\authorinfo{David M. Rogers}
{School of Natural Sciences and Mathematics,
University of South Florida, Tampa, Florida.}
{Email: davidrogers@usf.edu}

\maketitle

\begin{abstract}
  We present a C-language implementation of the lambda-pi calculus
by extending the (call-by-need) stack machine of Ariola, Chang
and Felleisen to hold types, using a typeless-
tagless- final interpreter strategy.
It has the advantage of expressing all operations
as folds over terms,
including by-need evaluation, recovery of the
initial syntax-tree encoding for any term,
and eliminating most garbage-collection tasks.
These are made possible by
a disciplined approach to handling the spine of each term,
along with a robust stack-based API.
Type inference is not covered in this work, but also derives
several advantages from the present stack transformation.
Timing and maximum stack space usage
results for executing benchmark problems are presented.
We discuss how the design choices for this interpreter
allow the language to be used as a high-level scripting language
for automatic distributed parallel execution of common
scientific computing workflows.
\end{abstract}

\section{ Introduction}

  Scientific computing workflows are commonly built at a high-level and require verifiability and reproducibility.
Although this is an ideal match for a functional programming style, almost all current
approaches are imperative.  In practice, these  quickly become special-purpose and non-composable
when dealing with large-scale projects.
Several frameworks have emerged to use task-level dependency information
to orchestrate data and work distribution, including HT-CONDOR\cite{dthai04},
Pegasus\cite{pegasus}, and many smaller projects such as REXEC\cite{bchun00} and
Fireworks\cite{fireworks}.
This also includes middle-ware libraries such as ADIOS\cite{adios} and StarPU.\cite{starpu}
One prominent example of what has been accomplished for a few specialized problems are the SETI@HOME and Folding@HOME projects.
Internally, many individual codes also uses dependency information to exploit fine-grained parallelism.
For example, the NAMD2 molecular dynamics\cite{lkale99} uses the Charm++ library,\cite{lkale13}
and large-scale multiphysics PDE-s solvers are moving in this direction as well.\cite{pnotz10}

  The idea that computer science can provide useful high-level abstractions
for scientific computing is well-established.\cite{sicm}
However, applications of functional language constructs in
high-performance and distributed computing
are just beginning to receive widespread recognition.  All three of
the Darpa high-productivity grand challenge languages,
X10, Fortress, and Chapel,\cite{hpcs}
included some language features for specifying programs
using their mathematical properties.
New languages like Tupeware\cite{tuple} are also emerging to facilitate
distributed workflows.

  All of these languages have been designed with user-directed parallelization in mind.
That design strategy targets parallelization of large, single-purpose, monolithic codes.
At the end of a run with such tools, all the data must be serialized, stored,
and then re-arranged for use with the next analysis.
For example, molecular dynamics with NAMD2 can be run on multiple, slightly different
molecular systems to generate millions of large arrays for later analysis.
Later, those sets of large arrays might be subject to map-reduce type computations,
as well as interactive analysis.  The entire workflow may later need to be re-run with slightly
different starting parameters or continued from the last run.  These glue steps almost always
require custom code, and consume a large portion of developer time in these workflows.

  {\em Our vision of a high-level language for distributed storage and execution of parsed data
requires  a new virtual machine that preserves term structure.}
The central objects in this workflow will not be files, but functional syntax trees.
Turning this picture into reality requires the ability to automate serialization of all data objects
and make retrieving and working with remote data transparent.
When required, single-purpose codes can be run as primitive operations,
executing on-demand as part of a global virtual filesystem / code ecosystem.
The evaluation method presented in this paper takes an important step by guaranteeing
that every evaluation intermediate can always be serialized, and that evaluation can stop at any point.
With this, fundamentally new applications are possible.

  We present details of an efficient, by-need evaluation strategy
on type-annotated terms that are capable of encoding the \LamPi{} calculus.
The bulk of the paper is devoted to introducing the stack form and proving
its equivalence with the usual, syntax tree encoding of lambda terms.
Reduction and normal terms are then defined in terms of the head-values
in stack form.  The equivalence between forms is directly proven by our implementation,
which provides functions converting back and forth between initial
and final representations.


  After detailing the term encoding and reduction machine, this paper describes
the fold structure based on the typeless- tagless- interpretation strategy,\cite{cjaqu09,okise10} including
all the details on garbage collection, refcounting, and handling of primitives.
Next, timings and space usage for two simple benchmarks are provided and compared
to the Glasgow Haskell Compiler
(GHC).\cite{ghc}  The implementation validates the claimed invariant properties of the stack-form,
and shows that the stack size remains manageable for even large problems.  We end by
discussing improvements that are possible, along with code parallelization strategies.

\paragraph{Prior Work}
  By-need reduction strategies have been extensively investigated as fully lazy evaluation
methods.\cite{zario95,zario97,jmara98}
More recently, Chang and Fellesein provided a new characterization of normal terms
and proofs of uniqueness, correctness, standard
reduction, and observational equivalence.\cite{schan12}  That work also introduced the single reduction
axiom, as well as a reduction machine.  Here, we fill in missing details in that picture
by re-stating their reduction axiom and proving invariants maintained by the stack representation.

  Part of the motivation for this work to find a better evaluation method for exploiting
implicit parallelism.  Parallel extensions to Haskell have been in-progress
for some time.\cite{eden}  This goal is also shared by recent innovations in
template\cite{tshea02} and cloud Haskell.\cite{jepst11}  Effective run-time code distribution
still remains a challenge for both of those frameworks.

\section{ Stack Representation}\label{s:stack}

\begin{figure*}
{\centering \begin{tabular}{rlll}
\multicolumn{3}{l}{Syntax} \\ \cline{1-2}
Stack [S],[T],[U] ::= & $\begin{bmatrix} &  \text{end} \\
                  &  \uparrow \\
                  \text{let} & x_1 : \text{[T] = [S]} | \text{(X)} \\
                     & \ldots \\
                     & x_n : \text{[T] = [S]} | \text{(X)} \\
                 \text{in} & \text{Val [S$_1$]} \cdots \text{[S$_m$]} \end{bmatrix}$ \\
                   \\
Values Val ::= & Var $x_i$ & Variable reference \\
                 $|$ & VarT $x_i$ & Type reference \\
                 $|$ & Ctor : [S] & Base data value / constructor (Int, Cons, ?, etc.) \\
                 $|$ & Dtor $x_i$ & Destructor \\
                 $|$& Prim : [S] & Primitive \\
\\
\multicolumn{3}{l}{Bottom Evaluation Semantics} \\ \cline{1-2}
Var $x_i$ & Done, & $x_i$ = Open or Var $x_i$ is recursive and WHNF$^1$ \\
Var $x_i$ & appendStack [S] [U] & eval($x_i$) = [U] \\
VarT $x_i$ & appendStack [S] [U] & eval($'x_i$) = [U] \\
Ctor : [S] & Done & \\
Dtor $x_i$ & appendStack [S] [U] & eval($x_i$) = Ctor [U] \\
Prim : [S] & Done & WHNF$^1$ or insufficient args \\
Prim : [S] & appendStack [S$_{k+1,\ldots}$] [U] & $\delta$ P$_k$ [S$_1$] $\cdots$ [S$_k$] = [U] \\
\end{tabular}\\}
\caption{Syntax, stack representation, and by-need reduction
         rules for a type-annotated $\lambda$-calculus.
         Sec.~\ref{s:props} details additional, essential restrictions on the locations of open binders, (X),
         and the ordering of end-pointers for each context.
         $^1$WHNF denotes that at least one open binder is reachable from the stack's context -- hence the
         term is already in weak head-normal form.
         The special notation in $\delta$-reduction for primitives notes that if a primitive
         consumes $k$ arguments, then those arguments are popped off of the pending applications
         when it returns.
        }\label{f:stack}
\end{figure*}

\begin{figure}
\includegraphics[width=0.5\textwidth]{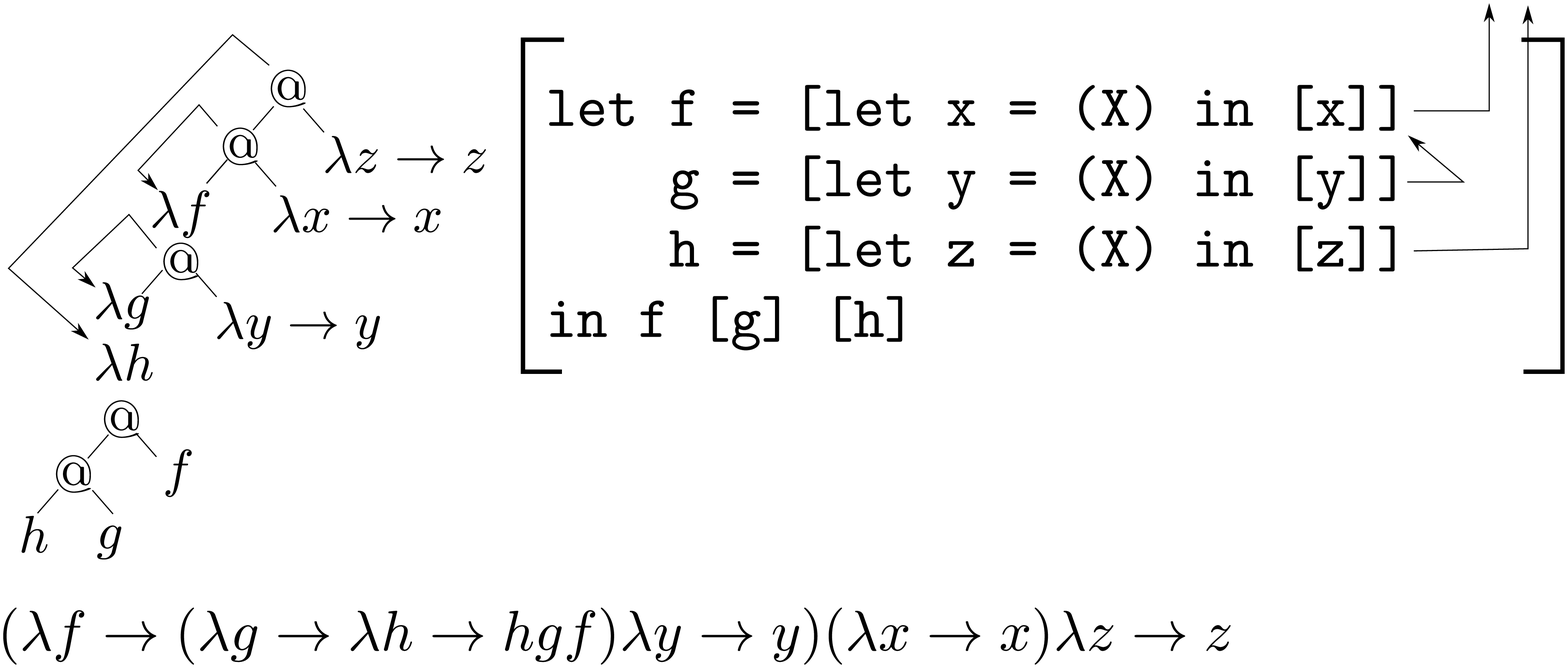}
\caption{Example reduction term, showing both initial and stack-based encodings.
Arrows on the left indicate pairing structure discovered during winding.  Arrows on the
far right show the end-pointers for the context of each bound right-hand side.
Each separate stack is delimited by $[]$-s.}\label{f:spine}
\end{figure}

  The interpreter uses a well-known transformation to turn every lambda term
into a single stack (Fig~\ref{f:wind}).  Each stack represents a spine for a lambda-term
(see Fig.~\ref{f:spine}).
In this form, the head-value is trivially available as the stack bottom (variable reference, constructor, etc.),
the list of all reachable variable bindings is available within the context,
and, until this stack is to be substituted somewhere else, all
the head-value's pending applications are visible as well.
Spines move to the left of applications and inside of lambda-s, and so
include a list of bound variables, as well as leftover, unmatched, lambdas and applications.
The critical information stored by our evaluator for each spine is listed in Fig.~\ref{f:stack}.

  The context is a series of type-annotated let-bindings
`owned' by the stack's closure.  They are used
for when needed by terms at the bottom of the stack,
and for determining which variables represent
function arguments.
The unpaired (open) binders are denoted by (X).  They have type annotations,
but no right-hand side.
There are no free variables in this representation.
A special rule prevents non-termination by dereferencing recursive bindings.
Variable references can point at either a term or a term's type
annotation.  Even though variable references are actually pointers to their bound value,
we carry the complete context to allow recovery of the initial syntax-tree encoding.

  Each context ends logically at the context where the stack branched
from a unique larger, enclosing, term.  These are shown schematically by arrows
in the example of Fig.~\ref{f:spine}.
Together, the end pointers have enough information to reconstruct
the pairing of @-$\lambda$-s in the initial, syntax tree, representation.

\begin{figure}
{\centering \begin{tabular}{rlll}
Initial Term & Next Step & Stack Action \\
\hline
Apply N M & N & pushAppl M \\
Lambda $x_i$ : N $\to$ M & M & pushCtxt $x_i$ : N = popAppl() \\
LetRec $x_i$ : N = L  in M & M & pushCtxt $x_i$ : N = L \\
Ctor : N & Done & Val = Ctor wind(N) \\
Dtor $n$ & Done & Val = Dtor $x_i$ \\
Prim : N & Done & Val = Prim wind(N) \\
Var $n$ & Done & Val = Var $x_i$ \\
VarT $n$ & Done & Val = VarT $x_i$ \\
\end{tabular}\\}
\caption{Specification of the winding operation, which
mutates a stack and terminates by setting the spine's head-value.
$N$, $M$ are abstract syntax trees (the initial encoding)
composed of the terms in the diagram.  Variable translation from de-Bruijn indices to context pointers, $x_i$,
occurs for the Var / VarT / Dtor rules by walking $n$ elements up the context for the current stack.
Open binders occur when no pending applications are present (i.e. when popAppl() is unsuccessful).}\label{f:wind}
\end{figure}

  This work aims to manipulate $\lambda$ terms entirely in stack form.  Accordingly, we
have defined stacks in Fig.~\ref{f:stack} so that the stack and syntax-tree
encodings are equivalent representations of the same object.  Our major contribution
is proving that a set of invariants guarantee this equivalence, and are preserved by
a sensible reduction strategy.  We then leverage this representation to define reduced terms
based on the stack definition, rather than the universally encountered initial definition of normal form.
We will also describe some other major advantages of working with the stack form.

  Many transformations that appear very difficult in the initial syntax tree representation
become almost trivial in stack form.
One of those advantages is the ability to make use of type annotations present for each binder.
In the \LamPi{} calculus, terms and types inhabit the same space.  The syntax definition in Fig.~\ref{f:stack}
makes this explicit, and at the same time loosens some of the conventional type-theory restrictions.
In particular, in \LamPi{}, the type of functions from values, $x$, of type $A$
to values of type $B(x)$ is ($\Pi x : A . B(x)$).  The $\Pi$ is a binding construct, since
$B$ can be an arbitrary function of $x$.  It is different from $\lambda$ only to denote that the whole term
must evaluate to a type-level quantity.  However, if $B(x)$ is another $\Pi$ function type,
then this restriction just means that the value returned by $B(x)$ is a type.  Therefore, $\Pi$ terms
can be curried in exactly the same way as functions.

  For simplicity, we choose to represent ,  $\Pi x:A . B(x)$ as just $\lambda x:A \to B(x)$.
The only difference from \LamPi{} is that a $\Pi$ value would have an earlier guarantee to be a type-level
quantity.  With this perspective, $\Pi$ is just a marking on a $\lambda$ value that guarantees it will
eventually return a type.  Finally, we include two special constructors for closing the type hierarchy.
A 0-argument hole constructor, ($? : N$),
to indicate an unknown term, and a 0-argument type-universe, ($\star : \star$).
In this way, type schemes can be translated to terms by applying a function to an unknown value.
For example, the type of the identity function, $\forall \alpha . \alpha \to \alpha$, can be written as
$(\lambda a : \star \to \lambda \_ : a \to a) (? : \star)$.  Applying this type to a right-hand side
actually shows that it is the typeof function, taking an argument to its type.

  We note that to be completely correct, $\star$ should be parameterized by
a universe level, or else it is possible
to run into a recursive definition problem known as Girard's paradox.  This universe level
is also closely connected to subtyping, which must distinguish $\star$ with unknown arity
from, e.g. (Int $: \star_0$).  Although preliminary results seem promising, this work will assume
all terms are correctly, statically typed, and will not address type inference issues.

  With this shortcut, it also becomes trivial to evaluate the type of every term that already has
properly annotated binders.
The type of each term can be found by typing the bottom of a stack, and then
dressing it with a new copy of all the binders in the stack's context.
For example, the type of the function,
\begin{equation*}
\lambda a : (?_1:\star) \to \lambda b : (?_2:\star) \to a\; b
\end{equation*}
requires solving the unification problem,
\begin{equation*}
(?_1 : \star) = \lambda \_ : (?_2 : \star) \to (?_3 : \star)
\end{equation*}
and results in the term,
\begin{align*}
\text{let t} &: \star = ?_2 : \star \\
\text{    u} &: \star = ?_3 : \star \\
\text{in  } & \lambda a : (\lambda (t:\star) \to u) \to b : t \to a\; b
\end{align*}

  The stack bottom's type is, in turn, just the type of the head-value,
after substitution of the head value for its type
and evaluation of the stack bottom.  For our example, this procedure results in,
\begin{align*}
\text{let t} &: \star = ?_2 : \star \\
\text{    u} &: \star = ?_3 : \star \\
\text{in  } & \lambda a : (\lambda (t:\star) \to u) \to b : t \to u
.
\end{align*}
Type-level holes are used for free type variables so that terms can represent type schemes
with the same ease as lambda.
This simplification should accordingly
make unification problems much clearer.  The type annotations on open binders must be unified,
followed by unification of the remaining context of one stack with the bottom of another.
For convenience, we also allow variables to reference
the type annotations on terms as well as the terms themselves.
This may be too permissive as it leads to multiple equivalent encodings -- with a type-level
hole present in the annotation or as a bound right-hand side.

\paragraph{ Example}

  Figure~\ref{f:spine} shows the stack winding transformation of an example from
Chang and Fellesein~\cite{schan12}.
The application right-hand sides are immediately apparent, but the difficulty comes in
accounting for the reference structure while evaluating the term.
All of these complications are stuffed into the appendStack method.

  Stacks are mutable structures that rearrange during appendStack.
First, the stack to be appended (the appendee) is copied.
Next, the appendor's head value is manually destroyed by calling stack\_dtor on
its internal type annotations, if present.
Finally, the appendee is inserted in the head position {\em between}
the last context value and the next pending application for the appendor.
Explicitly,
 \[
\boxed{\text{appendStack([S], [U]) = gc\_val([S]); wind([S], get\_ast([U]))}} \label{e:append}
 \]
Get\_ast retrieves the initial encoding of [U] without modifying [U].
The wind step traverses that syntax tree, pairing S's contexts when it finds an open
$\lambda$ and pushing its pending applications in front of the appendor's.

  In practice, get\_ast uses automatically garbage-collected heap-space
which is immediately orphaned, and so we have implemented appendStack as a direct copy instead.
For a direct copy, only the variable references (to contexts) internal to the copied stack need to be renumbered.
This is because variable references are direct pointers instead of de-Bruijn indices.
The appendee's entire context extends the appendor's.  Winding would have paired [U]'s
closed binders internally in exactly the same way, while its open lambdas
bind the appendor's pending applications.
The appendee's remaining pending applications go in front of the appendor's.

  In our implementation, stack deallocation (stack\_dtor), construction of the initial encoding (get\_ast),
and even by-need evaluation (need), are implemented using folds over the stack structure.
These folds are described in more detail in the next section.

\section{ Properties}\label{s:props}

  The context provides a `handle' for traversing the stack.
This is shown by noting that the stack maintains the all of the following invariants:
\begin{enumerate}
\item All sub-stacks directly referenced from a stack (annotations, bound right-hand sides and
pending applications) end at a context within the stack or at the stack's end.
\item All variables referenced from the stack are reachable on the path from
the stack's context pointer (but may be past the stack's end).
\item A sub-stack of the head value (type annotation or even a nested data structure not shown in Fig.~\ref{f:stack}) always ends at the start of the head value's context.
\item The ending points of bound right-hand sides in a given stack show
the pairing structure of lambda-s and applications in the following way.
If, when traversing the context from start to end, a right-hand side's end pointer
is another context value in the stack, no end pointers (of right-hand sides) on intervening contexts
can go beyond that context value.
Letrec-s are easily identified because their context ends at their own binder.
\item Pending application contexts end at ordered locations, with the
innermost, first, application closest to the start (having the largest context),
and the outermost, last, application closest to the end (smallest context).
\item No pending applications have end points between paired binders.
\item All open binders in a stack are reachable from all pending applications
for that stack.
\end{enumerate}
The first means that each referenced spine terminates at some
point of the main spine's context.  These termination points
are what hold the stack into a coherent tree -- since each stack represents
an entire self-contained sub-tree.  Nevertheless, the number of stacks
are exactly the same as the number of non-lambda or apply nodes in an initial
syntax tree encoding, while the number of contexts is exactly the number of lambda terms.

  The ordering properties are a consequence of the fact that each
spine represents a transformed syntax tree.  Application of a lambda
turns into a paired, closed, binding on the context, and so open binders have
to come before unpaired, pending applications.
Similarly, a pending application cannot have a context ending in-between
paired apply-lambda-s.  Pending applications are ordered
from outermost to innermost.

  We prove that as long as these invariants are satisfied, there
is always a transformation from the stack form back to a syntax tree.
Conversely, the translation from the syntax tree to the stack form
is trivial.
Because of this duality, the two forms are equivalent.  Operations
can then be specified on whichever form is more convenient.

  We then give an evaluation algorithm that simultaneously
computes both terms and types.  We prove this algorithm maintains
the above invariants.  Normalization is defined implicitly
via reducible and irreducible head-values.  This leads to
a computationally useful definition, since every stack
has a unique head-value, and since the reduction
proceeds by case analysis on these values.

They ensure that de-Bruijn indices can always be constructed by
counting the number of steps to a reference in the context.\cite{cmcb04}

  We now prove two theorems that establish the 1:1 mapping between
conventional lambda terms and stacks, and a third that justifies
defining normal forms in terms of a stack form.

\subsection{ Winding produces valid stacks}\footnote{As long as all variables are bound.}

  First, we assume that winding proceeds from an otherwise valid stack.
Since winding terminates by setting a bottom value, the starting bottom value
can be undefined.  We can always start at a stack with no contexts or pending applications.

  Since the winding process always proceeds along the left branch, we can consider
the set of $\lambda$,@, encountered as a series of tokens with unknown order.
We need to show that any series produces a valid stack.  The end-context of an application
is the nearest $\lambda$ to its left, while the applications always pair with the nearest $\lambda$
to their right.  Unmatched binders become open contexts, while unmatched applications
become pending applications.

  Property 1 is simple to show, since the context can only grow during winding, and originally starts at the stack's end point.  The context at every point during winding contains all binding $\lambda$-s in the tree's path to the root.  Violating property 2 would therefore mean that an unbound variable had been found.
Similarly, property 3 is automatically satisfied by all the winding termination rules, since the stack's context
is fixed when winding terminates.
  
  Properties 6-7 are also simple, since violating (6) would indicating a mis-pairing of $\lambda @ @$,
and since an unmatched binder cannot occur after any unmatched application.
Property 4 and 5 can be proven by noting the matching process is equivalent to matching parentheses.
Each @ acts as a left-parentheses, and each $\lambda$ as a right.  When traversing the
stack from start to end, it is like reading a parenthesized expression backwards.  The ending points
of the bound right-hand sides are everything left of its matched parentheses.  Property 4 follows
trivially from this observation.  Similarly, there may exist interspersed, unmatched $\lambda$-s and @-s.
These are always ordered.
  
\subsection{ Valid stacks can always be unwound to initial terms}\label{getast}

  We prove this by giving the unwind algorithm in Fig.~\ref{f:unwind}.  It proceeds in two phases.
First, the stack bottom is transformed into its initial encoding.  For stacks without annotations
this is trivial.  For stacks with annotations, the annotation must be unwound first.
Variables are simple to resolve to de-Bruijn notation by walking up the context (property 2).

  Second, the bottom is wrapped in a series of lambda/apply-s by visiting the context
from start to end (end is not visited).  Initially, the contexts of all pending applications are
copied, in the same order, onto an interrupt stack.
Unwinding stops at each interrupt to wrap the term in an Apply
(unwinding the pending application right-hand side).
At each binder, the term is wrapped in a Lambda tree node (or a LetRec if the right-hand side context ends at the current binder).
Using the analogy from the previous section, each binder in the context is a right-parentheses.
The context of its corresponding left-parentheses is pushed onto the interrupt stack
and unwinding proceeds to it.  No pending application can interrupt this process (by property 6).
At this point, the term is wrapped in an Apply node.
This must match the corresponding Lambda, since all other Lambda-s will have been matched
using this procedure, by property 4.

  Interrupts are guaranteed to be encountered during unwinding by property 1.
If unwinding encounters an open binder, no interrupt is pushed.  The interrupt
stack will already be clear, by property 7.

  We also observe here that each paired application can be destroyed without
spoiling the stack property.  By analogy to the parentheses matching, none of the ordering
properties are altered.  However, this involves checking the end-pointers of every stack reachable
from the current one that ends at the collected binding, replacing it with the next binder in the series.
This can cause a considerable slow-down and should be re-examined in the future.
Our implementation maintains reference counts for each binder to identify and remove
unused binders immediately.

\subsection{ Evaluation does not break validity}

  Since evaluation makes use of essentially only 1 rule, this is easy to show.
First, note that appendStack can be written in terms of wind and unwind (Eq.~\ref{e:append}).
Destroying the stack bottom is inconsequential, and results in a valid starting stack for wind.
Winding the result of an evaluation may encounter any variables reachable from that evaluation.
Since only right-hand sides are substituted, those variable references fall in one of two sets.
Either the variable is internal to the right-hand side (below the right-hand side's end pointer),
or it is external (at or above end).  If internal, that reference will link to a binder that will be added during wind.  By property 1, that end pointer is reachable from the current stack's context.
Therefore, the precondition for wind to produce a valid stack has been met.

  Evaluation can trivially stop at any point and still produce a valid partial evaluation.

  Next, we should consider the special impact of the primitive reduction rule.
In our implementation, primitives use an API for manipulating the first $k$
terms in the pending application stack (where $k$ depends on the primitive).
These are evaluated by-need when the primitive requests their value.
The API makes it impossible to insert binders or change context values
(other than normal evaluation).
The primitive completes by leaving a return value on the application stack,
and destroying the others.  This last remaining value forms the new
stack bottom.
To show this is valid, note that none of the ordering properties are violated
if any $k-1$ of the first $k$ pending applications are removed, since no re-ordering
is possible.  Also, the move of the remaining stack to the bottom also uses appendStack.

  If more generality is needed, the $k$ pending applications could
all be re-linked to end at the start of the stack's context and a general lambda-term
could be wound into the head value.
The additional invariants have been very useful for developing and testing
functions that work with stacks, including fast copies and the primitive API.

\section{ Unwinding Functions}

  Dealing with in-memory references is notoriously difficult in low-level
languages.  This section presents wind and unwind routines that
standardize computations, eliminating most common errors.
The major workhorse is the unwind function, which essentially works
as the foldr of {\tt f}, starting from {\tt val}.  In higher-level languages,
it is possible to achieve a much more precise manipulation of the function,
including composition, partial evaluation, and inlining.  These three together
produce the tagless property of tagless interpreters.

  Underneath the hood, each member of a deconstructor typeclass has to be looked
up from a dictionary and replaced with the appropriate function during compilation
of the evaluator.  Here, we make that choice explicit by providing different definitions
of val, let\_l, let\_a, and apply for each type of fold.

\begin{SaveVerbatim}{f:unwind}
struct SFold {
    void *(*val)(struct Program *p, struct State *s, void *val);
    void *(*let_l)(struct Program *p, struct Ctxt *c, void *val);
    void *(*let_a)(struct Program *p, void *val, struct State *b);
    void *(*apply)(struct Program *p, void *a, struct State *b);
};

void *unwind(struct Program *p, struct State *s,
                      struct SFold *f, void *val) {
    struct State *rhs, *next, *sup;
    struct Ctxt *c;
    void *ret;

    while( (ret = f->val(p, s, val)) == NULL);
    val = ret;  // Caution! f->val() may deref Var-s,
    sup = s->up; // and change `s->c, s->up'
    c = s->c;

    while(sup != NULL) {
        next = sup->parent;
        val = unwind_context(p, &c, sup->end, f, val);
        val = f->apply(p, val, sup); // ascend application
        sup = next;
    }

    return unwind_context(p, &c, s->end, f, val);
}

void *unwind_context(struct Program *p, struct Ctxt **cp,
        struct Ctxt *end, struct SFold *f, void *val) {
    struct State *sb;
    struct Ctxt *next, *bend, *c = *cp;

    while(c != end) {
        sb = c->b; // just in case c is freed (e.g. by let_l)
        next = c->next;
        if(sb != NULL)
            bend = sb->end; // or end is modified...

        val = f->let_l(p, c, val);

        if(sb != NULL && bend != c) { // unwind sub-contexts
            val = unwind_context(p, &next, bend, f, val);
            val = f->let_a(p, val, sb);
        }
        c = next;
    }
    *cp = c;
    return val;
}
\end{SaveVerbatim}
\codefig
{f:unwind}
{Stack-unwinding algorithm, expressing a generic fold.  This is the heart of the interpreter.}

\paragraph{ Recovering Initial Encoding (get\_ast)}
  The visitor structure in our implementation is specially designed for this task.
It can produce arbitrary partial terms and context errors by recognizing the end pointer of its starting stack
and jumping in number to a root context, which is visible from another arbitrary place in the term.
In pseudocode,
\begin{itemize}
\item val (StackVar cp) = val $\gets$ Var (steps from s->c to cp)
\item  let\_l x t =  val $\gets$ Lambda x : get\_ast(t) $\to$ val
\item let\_a b = val $\gets$ Apply val get\_ast(b)
\item apply b = val $\gets$ Apply val get\_ast(b)
\end{itemize}
A major source of early difficulties was attempting
to both evaluate and produce a syntax tree at once.

\paragraph{ Call-By-Need}

   Call-by-need is focused on evaluating the stack's head-value, but also has the ability to clean up the context or evaluate pending applications after it is completed.  How to evaluate pending applications is one major point of contention in existing applications.\cite{smarl04}  If the stack is to be substituted, or if the head value is strict, evaluation needs to be done.  However, this work can also be delayed by creating new bindings on the context.  This requires a generalized lifting operation.

\begin{itemize}
\item val: Implement Fig.~\ref{f:stack}, returning NULL after appendStack and returning the current stack on Done.
\item  let\_l Ctxt c $|$ nref c == 0 = stack\_dtor c.t; stack\_dtor c.b; unlink c
\item  let\_l Ctxt c $|$ otherwise = need (c.t)
\item let\_a b = pass
\item apply b = need b
\end{itemize}

  xx has clearly described how call-by-need can be implemented in terms
of folds over this structure.  Whenever a head-value is a variable reference,
the binding for that variable is evaluated by-need, and the result
is copied and substituted for the variable.  Because all variable references
are pointers in this scheme, no index re-numbering is required during the copy.

  Substitution can often be optimized to a direct
move of the stack.  This is possible when the variable is the only reference
to its binder.  We maintain reference counts by incrementing a context's
count whenever a variable is added during {\tt wind},
and decrementing reference counts whenever a variable is destroyed.
Recursive references can not be optimized in this way,
since substitution might lead to more references.
Recursive references are only counted for references outside the stack.
Cyclic references require a special constructor (we use record types).
Their counts are maintained by saving the reference
count before and restoring it after their right-hand
side is evaluated.

  As shown by the use of stack\_dtor, by-need evaluation proceeds
in tandem with stack destruction.
Further traversal steps continue up the tree, visiting all
the binders in the context.
After the stack bottom has been evaluated, all the needed
right-hand sides should have been substituted.  Hence,
all matched let-bindings could destroyed, leaving only the open
let-bindings.  A refinement used in this work is to keep
all bindings that have any references remaining.  This way,
exactly as much garbage collection is done as needed
-- so partial evaluation still results in a valid term.
The correctness of this procedure relies critically on the
fact that the tree is traversed from start to end.  Because
of this, each let-binding is visited immediately after all possible
references to it have already been visited.

\paragraph{ Garbage Collection (stack\_dtor)}

  Destroying the stack can be done by de-allocating all its internal data structures.
It only makes sense to do this with a visiting fold.  We have taken special care
to read the next few steps from the stack during the unfolding process
so that this works.  Since stacks maintain a strict tree ordering,
automatic garbage-collection is not needed (or used in our implementation)
for stack structures.

\section{ Benchmarks}

  We give timings and heap usage below for the n-queens and
tak benchmark on an Intel 2.5 GHz Core I7 running OSX, for an
interpreter compiled with clang-602.0.53.  Each test was checked for reproducibility
and status was tabulated at each context creation.  Timings were
reported for the non-heap reporting versions.

\begin{figure}
\includegraphics[angle=-90,width=0.5\textwidth]{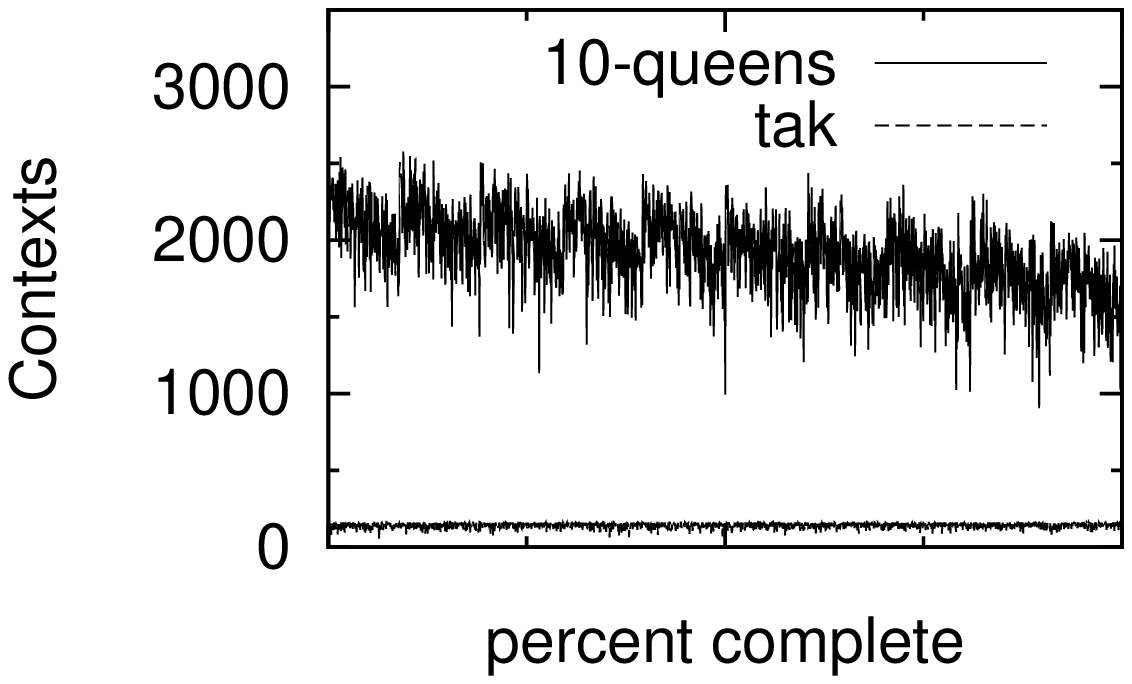}
\caption{Both benchmark programs are executed by our interpreter in constant stack-space,
as shown by contexts vs. run-time.
In both cases, the number of stacks (and hence total running size)
is proportional to the number of contexts.
Corresponding sizes are given in the text -- 738 kbytes for 10-queens
and 77 kbytes for tak.}\label{f:use}
\end{figure}

  The 8-queens program runs in 6.9 seconds and makes 1.8 million
context allocations, with a high-water mark
of 2029 contexts and 7856 stacks.
The 10-queens program runs in 221 s and makes 45 million context
allocations, with a high-water mark of 2778 contexts and 10789 stacks (Fig.~\ref{f:use}).
The program utilized the deforested version of lists, with type
\begin{equation*}
\text{List} (a : \star) = (\lambda b:\star \to (\_:a \to \_:b \to b) \to \_:b \to b) (?:\star)
\end{equation*}
introduced by Gill, Launchbury, and Peyton-Jones.\cite{agill93}
Interestingly, that strategy is an incipient version of the tagless method -- as it
relies on replacing the list construction operators, cons and nil,
with deconstruction functions.

  The tak benchmark is a stringent test of recursive calls.
When run with our interpreter on a medium problem size (27 16 8),
it makes 26.9 million context allocations.  Nevertheless, its high-water mark
was just 162 contexts and 1245 stacks (Fig.~\ref{f:use}).
The benchmark ran in 119 s without any memoization.

  In contrast, the Glasgow Haskell Compiler (GHC)\cite{ghc} uses highly optimized array STM and
pinned memory to solve the 10-queens problem in just over 0.01 s,
and tak in 0.09 s.
For both problems, its heap-space usage remains essentially flat at 36 kbytes throughout the run.
It should be noted that GHC has much slower performance for these problems in interpreted
execution mode, requiring 1.4 s to run 10-queens and 4.9 s to run tak.

  Our linked-list implementation of stacks and contexts requires 48 bytes per context and
56 bytes per stack.  This makes the high-water marks for 10-queens at 738 kbytes
and tak at 77 kbytes.  These compare favorably with GHC's heap-space usage.

\section{ Discussion}

  The benchmark problems above clearly show that the interpretation strategy taken here
is robust.  Development versions of our interpreter have successfully checked all the invariants
listed in Sec.~\ref{s:prop} at every major step of code interpretation.
  Although the run-times above are slightly disappointing, it is highly encouraging that
the stack space usage parallels that used by GHC itself.  It is notable that most naive functional
language implementations face great difficulty with these programs.

  This is an excellent starting point for more advanced evaluators
for the genereric lambda-calculus with dependent types.
Several improvements can likely be made to increase the execution speed,
and to finally realize the automatic task-level parallelism that $\lambda$-calculus
has the potential to provide.

   The actual implementation recognizes Cons, Tup, and Commit (for a directory of named bindings),
 as constructors, provides several base data types, as well as provides a means for creating
 generalized algebraic datatypes\cite{mzeng01} as Sym constructors.  It implements IO by
 recognizing stacks with IO head-values as constructors during functional evaluation, and as primitives
 during imperative reduction.\cite{pwadl97}  IO primitives return an extra flag indicating whether evaluation
 has completed with returnIO.  The ST monad is implemented in a similar way, but can be invoked
 from within a functional computation by the runST primitive,
 which creates the initial copy and manages the state.
 It also provides a unique mechanism for creating primitives and new packed binary data values
 from specially formatted Commit types.
  
\paragraph{ Elimination of most relinking steps}

  As mentioned in the text describing the stack\_dtor fold, a major source of inefficiency in the present work
is the maintenaince of end pointers on each context.  Since evaluation proceeds
in tandem with context deletion, the stack must be constantly scanned to rewrite
these pointers.  If, instead, the end pointer of every right-hand side was contained
as part of its context, then each stack could have a NULL end-pointer.

  This would eliminate recursive re-writes, but require some other mechanism
for creating de-Bruijn indices pointing outside the current stack during get\_ast.
One simple resolution is to track parent spines for each spine.
Alternatively, the parent spine of each context could be used.
The spine tree structure could be re-built by
examining all Var pointers in each stack and creating a topological sort.

\paragraph{ Unboxing and inlining fold functions}

  The usual machine implementation of the lambda-calculus uses continuation passing
to jump back after evaluating an application right-hand side.  The unwind method
is slightly more complicated, maintaining a state of paired binders.  Nevertheless,
it could be made into a closure and returned into in the same way.

  Once this step has been taken, each fold function over stacks can be inlined
by turning all stacks into executable code.  For example, the
code for {\tt var $\gets$ Apply(var, get\_ast(b))} would be push\_closure (Apply(var, . ), closure)
followed by call b.  For arguments that have statically known, machine-representable types,
both the producer and consumer know can create code that makes use
of unboxed data values -- final form.

\paragraph{ Enhanced Type System Development}

  As discussed in Sec.~\ref{s:stack}, terms and types are syntactically very close.
This makes unification much simpler, since unifying two terms in stack
form just requires unifying the types of their open binders, and then unifying
the bottom of the term with the smaller context with the remaining stack
on the other.  It is the unification solutions which make this process difficult.
Each solution eventually gets written into a ? constructor.  However, that
solution must be lifted to a point in the context where it is reachable from both halfs of the
unification problem.  If the solution references variables, those must also be lifted.
This reachability requirement translates to an occurs-check so that the solution
cannot be recursive.  It also dis-allows open binders (or stacks referencing open binders)
from being lifted above the binder.  This imposes a very helpful restriction on type annotations,
namely that unification cannot automatically create types that depend on future, unknown variables.
This requires a non-trivial edit in stack form which will be discussed in future work.  

\paragraph{ Distributed Parallel Evaluation}

  The interpreter built in this work can serialize syntax trees in initial form to a Google protocol buffer format, and store them in a distributed hash table.  The resulting hash table manages objects in much the same way as a git source-code repository.  This automatic serialization of every term
 is made possible by the tagless-inspired strategy of this work.  From this point, trivial parallelization
 of arbitrary code can be facilitated using a bag of tasks model.

  We note that a distributed, lazy, work-stealing scheduling strategy is also possible.
If a term has a strict head-value, all its pending application stacks will need to be evaluated.
By properties 1 and 2, all of the references from any of these applications
are reachable from the context of the current stack.  Therefore, only the bound right-hand sides
need to be made available in case they are needed by a peer evaluating of one of those
pending applications.  Once placed in the table, the context can be replaced with a hash reference.

  The work list starts out when each strict, pending application is made available as code
{\em and} placed in an active queue.  In general, peers can start by dequeuing any object
from the active queue, and activating any right-hand sides as needed.
Peers remain lazy, evaluating references only by-need.
On completion, the evaluated term is stored in a completed work list, along with
a mapping from the hash of the original to the hash of the evaluated term.
Generally, peers may cause contention by competing to dequeue common references.
Nevertheless, re-evaluation does not cause harm, and this organization structure
gives a implementable distributed lazy evaluator.


\begin{small}

\appendix

\end{small}
\end{document}